\documentclass[aps,prc,twocolumn,superscriptaddress,showpacs]{revtex4}
\usepackage{graphicx,color}

\newcommand{\defeq}{\mathrel{\mathop:}=}
\usepackage{amsmath}

\begin{document}

\title{\textit{Ab initio} effective interactions for $sd$-shell valence nucleons}  

\author{E. Dikmen}
\email[]{erdaldikmen@sdu.edu.tr}
\affiliation{Department of Physics, Suleyman Demirel University, Isparta, Turkey}
\affiliation{Department of Physics, University of Arizona, Tucson, Arizona 85721}
\author{A. F. Lisetskiy}
\email[]{lisetsky@comcast.net}
\altaffiliation{currently at Mintec, Inc., Tucson, Arizona}
\affiliation{Department of Physics, University of Arizona, Tucson, Arizona 85721}
\author{B. R. Barrett}
\email[]{bbarrett@physics.arizona.edu}
\affiliation{Department of Physics, University of Arizona, Tucson, Arizona 85721}
\author{P. Maris}
\email[]{pmaris@iastate.edu}
\affiliation{Department of Physics and Astronomy, Iowa State University, Ames,
Iowa 50011}
\author{A. M. Shirokov}
\email[]{shirokov@nucl-th.sinp.msu.ru}
\affiliation{Department of Physics and Astronomy, Iowa State University, Ames,
Iowa 50011}
\affiliation{Skobeltsyn Institute of Nuclear Physics, Lomonosov Moscow State University, 
Moscow 119991, Russia}
\affiliation{Pacific National University, 136 Tikhookeanskaya st., Khabarovsk 680035, Russia}
\author{J. P. Vary}
\email[]{jvary@iastate.edu}
\affiliation{Department of Physics and Astronomy, Iowa State University, Ames, 
Iowa 50011}

\date{\today}

\begin{abstract}
We perform \textit{ab initio} no-core shell-model calculations for $A=18$ and $19$ 
nuclei in a $4\hbar\Omega$, or $N_{\rm max}=4$, model space by using the effective JISP16 
and chiral N3LO nucleon-nucleon potentials and transform the many-body effective Hamiltonians 
into the $0\hbar\Omega$ model space to construct the $A$-body effective Hamiltonians 
in the $sd$-shell. We separate the $A$-body effective Hamiltonians with $A=18$ and 
$A=19$ into inert core, one-, and two-body components. Then, we use these core, one-, and 
two-body components to perform standard shell-model calculations for the 
$A=18$ and $A=19$ systems with valence nucleons restricted to the $sd$ shell. 
Finally, we compare the standard shell-model results in the $0\hbar\Omega$ model space
with the exact no-core shell model results in the $4\hbar\Omega$ model space 
for the $A=18$ and $A=19$ systems and find good agreement.
\end{abstract}


\maketitle

\section{Introduction}

In recent years remarkable progress in \textit{ab initio} microscopic 
nuclear structure studies has been made in calculating nuclear properties, 
e.g., low-lying spectra, transition strengths, etc., in light nuclei. 
Large-basis \textit{ab initio} no-core shell-model (NCSM) calculations, 
which provide the foundation for this investigation, 
have been successful in reproducing the low-lying spectra 
and other properties of nuclei with $A\le16$
\cite{Stetcu_2005,Navratil_2005,Stetcu_2006,Nogga_2006,Roth_2007,
Maris_2009,Maris_2010_1,
Maris:2011as,Cockrell_2012,Maris_2013,
Navratil_1997,Navratil_2000_1,Navratil_2000_2,Navratil_2001,
Navratil_2007,Barrett_2013,Jurgenson_2013,Shirokov_2014,Maris_2014}. 

In NCSM calculations all nucleons in the nucleus are active and treated 
equivalently in the chosen model space. 
When we increase the model space to obtain more precise results, we encounter
the problem that the size of the calculations can easily exceed currently 
available computational resources. This is especially true as one proceeds 
towards the upper end of $p$-shell nuclei and beyond.
The problem may be cast as a challenge to reproduce the many-body correlations
present in the large space in a tractable, smaller model space. Success in
this endeavor will open up the prospects for \textit{ab initio} solutions 
for a wider range of nuclei than are currently accessible. 

The NCSM has proven to be an \textit{ab initio} microscopic nuclear structure
approach that has been able to reproduce experimental results and to make
reliable predictions for nuclei with $A \leq 16$. These successes
motivate us to develop approaches for heavier-mass nuclei. In one approach, 
a small model space effective interaction has been constructed
by modifying the one-body piece of the effective two-body Hamiltonian and 
employing a unitary transformation in order to account for many-body 
correlations for the $A$-body system in a large space \cite{Fujii_2007}.
In another approach~\cite{Lisetskiy_2008}, the effective two- and three-body 
Hamiltonians for $p$-shell nuclei have been constructed by performing 
$12\hbar\Omega$ \textit{ab initio} [i.e., $N_{\rm max}=12$ harmonic oscillator 
(HO) quanta above the minimum required]
NCSM calculations for $A=6$ and $A=7$ systems and explicitly projecting the many-body 
Hamiltonians onto the $0\hbar\Omega$ space. 
These $A$-dependent effective Hamiltonians
can be separated into core, one-body, and two-body (and three-body) components, 
all of which are also $A$-dependent~\cite{Lisetskiy_2008}. 

Recently, two more \textit{ab initio}
methods for valence nucleon effective interactions have been introduced with 
the same goals; one is based on the in-medium similarity renormalization group 
approach \cite{Bogner_2014} and the other is based on the coupled-cluster 
method \cite{Jansen_2014}.

In this work, following the original idea of Refs. \cite{Navratil_1997,Lisetskiy_2008},
we derive two-body effective interactions for the $sd$ shell by using 
$4\hbar\Omega$ NCSM wave functions at the two-body cluster level, which contain all
the many-body correlations of the $4\hbar\Omega$ no-core model space. 
The goal of this work is to demonstrate feasibility of this approach in the 
$sd$ shell, where we do not require calculations at the limit of currently accessible 
computers. Such a major extension will be addressed in a future effort.

\begin{figure}[t]
\includegraphics[width = \columnwidth]{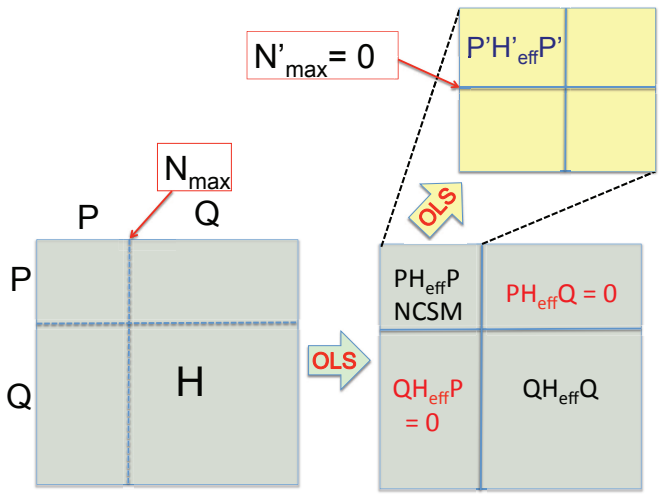}%
\caption{(Color online) Flow of renormalizations adopted to derive an effective interaction 
for valence nucleons. The OLS procedure is first applied to derive a NCSM effective 
interaction for the full $A$-nucleon system resulting in the ``primary'' effective Hamiltonian 
$PH_{\rm eff}P$ for the chosen no-core basis space (the ``$P$-space'') indicated on the large 
square on the right of the figure in its upper left corner. The many-body truncation is 
indicated by $N_{\rm max}$, the total number of HO quanta above the minimum for that system. 
The OLS procedure is applied again by using the results of the NCSM calculation to derive the 
``secondary'' effective Hamiltonian $P'H'_{\rm eff}P'$ for the chosen valence space 
(the $P'$-space with the smaller many-body cutoff $N'_{\rm max}$) indicated on the square 
in the upper right of the figure.}
\label{Figure_Double_OLS}
\end{figure}

At the first step, we construct a ``primary'' effective 
Hamiltonian  following the Okubo-Lee-Suzuki (OLS) 
unitary transformation method \cite{Okubo_1954,Suzuki_1980,Suzuki_1982}. 
We indicate this first step schematically by the progression shown with 
the two large squares in the lower section of Fig.~\ref{Figure_Double_OLS}.
We elect to perform this first step at the two-body cluster level 
for $^{18}$F in the $4\hbar\Omega$ model space (the ``$P$-space'')
following the NCSM prescription \cite{Navratil_2000_1,Navratil_2000_2,Barrett_2013}.
For our initial interactions we select the JISP16 \cite{Shirokov_2007} 
and chiral N3LO \cite{Entem_2003} potentials.
Our formalism may be directly adapted to include the three-nucleon force (3NF) 
but the computational effort increases dramatically. Thus, we do not include 
the 3NF in this initial work.

For the second step, we begin by performing a NCSM calculation for $^{18}$F with the primary 
effective Hamiltonian in the $4\hbar\Omega$ model space to generate the low-lying eigenvalues 
and eigenvectors needed for a second OLS transformation as indicated by the flow to the upper 
right in  Fig.~\ref{Figure_Double_OLS}. These $^{18}$F eigenvectors are dominated by 
configurations with an $^{16}$O system in the lowest available HO orbits and two nucleons in 
the $sd$ shell.  All additional many-body correlations are also present.  With these $^{18}$F 
eigenvectors and eigenvalues we then solve for the ``secondary'' effective Hamiltonian, again 
using an OLS transformation, that acts only in the $N'_{\rm max} = 0$ space of $^{18}$F but 
produces the same low-lying eigenvalues.  Here we are following the scheme initially introduced 
in Ref. \cite{Navratil_1997}. The matrix elements of this secondary effective Hamiltonian have 
the property that all configurations are defined with two nucleons in the $sd$-space and an 
$^{16}$O subsystem restricted to the lowest available HO single-particle states. This second 
step therefore produces a secondary effective Hamiltonian that is equivalent to what we would 
call the 18-body cluster Hamiltonian in the NCSM acting in the $N'_{\rm max} = 0$ space.

At the third step, we carry out NCSM calculations for the $^{16}$O, $^{17}$O, and  $^{17}$F
systems with the primary effective interaction in the $4\hbar\Omega$ basis space. The results 
of these calculations produce, respectively, the core and one-body components included in the 
secondary effective Hamitonian.

At the fourth step, we subtract the core and one-body terms from the  secondary 
effective Hamiltonians of step 2 to obtain the effective valence interaction
two-body matrix elements (TBMEs) in the $sd$-shell space. 

Following the completion of these four steps, we then use the effective valence 
interaction matrix elements along with the extracted single-particle energies 
(for both the proton and the neutron) for standard shell-model (SSM)
calculations in the $sd$-shell space. 

For any system with $A > 18$, we can obtain its 18-body cluster Hamiltonian by 
repeating the entire procedure utilizing the primary effective Hamiltonian for 
that value of $A$.  The second and subsequent steps remain the same. That is, 
we perform NCSM calculations with the primary ($A$-dependent) effective $NN$ potential 
for $^{16}$O, $^{17}$O, $^{17}$F, and $^{18}$F in order to obtain the ($A$-dependent) 
core energy, single-particle energies, and TBMEs, which can then be used in a SSM 
calculation for that value of $A$. We provide details for applications to  $A > 18$ 
systems below using $^{19}$F as an example.

We employ the Coulomb interaction between the protons in the NCSM calculations 
which gives rise to the major shift between the derived neutron and proton 
single-particle energies. Exploration of full charge-dependence in the derived 
two-body valence interactions will be addressed in a future effort. In particular, 
our current $A=18$ and $19$ applications will have at most one valence proton 
so we do not require a residual Coulomb interaction between valence protons in 
this work. 

For the chiral N3LO we retain full charge dependence in the first step that is, when 
deriving the primary effective Hamiltonian. Thus, the $A$-body, core, and valence system 
calculations are performed with full charge dependence retained.  Since we currently solve 
only for $^{18}$F in step 2, we derive only the isospin-dependent but charge-independent 
secondary effective Hamiltonian. To retain full charge dependence in the secondary 
effective Hamiltonian, which would constitute predictions beyond conventional 
phenomenological interactions, would require additional $^{18}$O and $^{18}$Ne 
calculations that are intended in future efforts.

One may straightforwardly generalize these steps outlined above to solve for effective 
three-body valence interactions suitable for SSM calculations.  Earlier efforts using 
an alternative implementation of step 3 ~\cite{Lisetskiy_2008} showed that effective 
three-body valence interactions lead to significant improvements over effective two-body 
valence interactions.


\section{Theoretical Description}

\subsection{No-core shell model and effective interaction}

The NCSM calculations start with the intrinsic Hamiltonian of the $A$-nucleon system,
omitting any 3NF in the present effort:
\begin{eqnarray}
\label{H_intrinsic}
H & = & \sum_{i<j=1}^{A}\frac{(\vec{p_i}-\vec{p_j})^2}{2Am}
        +\sum_{i<j=1}^{A}V_{ij}^{NN} \nonumber\\
  & = & T_{\rm rel} + V^{NN}, 
\end{eqnarray}
where $m$ is the nucleon mass, $V_{ij}^{NN}$ is the bare $NN$ interaction, $T_{\rm rel}$ 
is the relative kinetic energy and $V^{NN}$ is the total two-body interaction. We will 
add the Coulomb interaction between the protons at a later stage since we treat it as 
a perturbative correction to the derived  primary effective Hamiltonian. In order to 
facilitate convergence, we modify Eq.~(\ref{H_intrinsic}) by adding  (and later subtracting) 
the center-of-mass HO Hamiltonian which  introduces a dependence on the HO energy, 
$\hbar\Omega$, and this dependence is denoted by ``$\Omega$'' in what follows. In addition, 
we introduce $a \leq A$ to define a new $a$-, $A$-, and $\Omega$-dependent Hamiltonian:
\begin{equation}
\label{H_A_body}
 H_{a}= \sum_{i=1}^{a}\left[\frac{\vec{p_i}^2}{2m} 
                        + \frac{1}{2}m\Omega^2\vec{r_i}^2\right]  
                        + \sum_{i<j=1}^{a}V_{ij}(\Omega,A),
\end{equation}
where $a=A$ corresponds to the full Hamiltonian of Eq.~(\ref{H_intrinsic}) with the center-of-mass 
HO Hamiltonian added and $V_{ij}(\Omega,A)$ is the modified bare $NN$ interaction which 
we define independent of the parameter $a$ but including dependence on $A$:
\begin{equation}
\label{V_Omega_A}
 V_{ij}(\Omega,A) = V_{ij}^{NN}-
                    \frac{m\Omega^2}{2A}(\vec{r}_i-\vec{r}_j)^2.
\end{equation}
The exact solution of  Eq.~(\ref{H_intrinsic}) for a subset of its eigensolutions in a 
finite matrix diagonalization requires the derivation of an $A$-body effective interaction 
for sufficiently heavy nuclei \cite{Barrett_2013}, but such a derivation is not currently 
possible for $A >5$ with realistic interactions.

Here, we adopt the two-body cluster approximation ($a=2$) for the effective interaction 
\cite{Navratil_2000_1,Navratil_2000_2}. This allows us to solve the eigenvalue problem for 
a sufficiently large basis space that we achieve convergence of a suitable set of low-lying 
eigenvalues and eigenvectors needed to construct the primary effective Hamiltonian.
In the $a=2$ approximation, the Hamiltonian (\ref{H_A_body}) becomes
\begin{equation}
\label{H_2_body}
 H_{2} = \sum_{i=1}^{2}\left[\frac{\vec{p_i}^2}{2m}
                    + \frac{1}{2}m\Omega^2\vec{r_i}^2\right] 
                    + V_{12}(\Omega,A).
\end{equation}
For deriving an effective three-nucleon interaction one would take $a=3$.
Note that the $A$-dependence enters the Hamiltonian $H_{2}$ through 
the second term in Eq.~(\ref{V_Omega_A}). For example, this $A$-dependence 
makes the two-body cluster Hamiltonian $H_{2}$ in the $T=0$ channel 
different from the deuteron Hamiltonian. 
In order to preserve Galilean invariance in the primary effective Hamiltonian, 
we obtain the solutions to Eq.~(\ref{H_2_body}) in the relative HO basis 
where the the center-of-mass component of the first term in Eq.~(\ref{H_2_body}) 
plays no role.

We now introduce our representation of the unitary transformation needed to 
construct the primary effective Hamiltonian $PH_{\rm eff}P \defeq H_{2}^{P}$ in the 
$P$ space (signified by a superscript ``$P$'') of the first step.
The $P$-space effective interactions have $A$ dependence, $\Omega$ dependence and 
$N_{\rm max}$ dependence all implied by the superscript $P$.
We define $N_{\rm max}$ as the maximum number of HO quanta in the many-body HO 
basis space (the NCSM basis space) above the minimum for the $A$-nucleon nucleus. 
We select $N_{\rm max}=4$ in the present work. The resulting finite $P$ space, 
of dimension $d_P$, for the first step is indicated on the left-hand side of 
Fig.~\ref{Figure_Double_OLS}. The diagonalization of the Hamiltonian $H_{2}$ in the 
relative HO basis provides the unitary transformation $U_{2}$ such that
\begin{equation}
\label{H_diag}
 H_{2;\rm diag}=U_{2}H_{2}U_{2}^{\dagger},
\end{equation}
where $H_{2;\rm diag}$ is the diagonal matrix containing the eigenvalues $E_{2;k}$:
\begin{eqnarray}
\label{H_diag_mf}
 H_{2;\rm diag}=\left(\begin{array}{cccc}
      E_{2;1} & 0         & \ldots & 0           \\
      0         & E_{2;2} & \ldots & 0           \\
      \ldots    & \ldots  & \ldots & \ldots      \\
      0         & 0         & 0   & E_{2;\rm max} \\
                       \end{array} \right), 
\end{eqnarray}
where the subscript ``$\rm max$'' signifies the dimension of the $a=2$ space sufficient 
to guarantee convergence of the $d_P$ low-lying eigenvalues and eigenvectors. We typically 
employ $\rm max = 200~{\rm to}~450$ for a realistic $NN$ interaction governed by the need 
to converge the results for the chosen interaction at the selected value of $\hbar\Omega$.

By introducing the model space $P$, one builds the matrix $H_{2;\rm diag}^{P}=PH_{2;\rm diag}P$:
\begin{eqnarray}
\label{H_diag_pmax}
 H_{2;\rm diag}^{P}=\left(\begin{array}{cccc}
      E_{2;1} & 0         & \ldots & 0               \\
      0         & E_{2;2} & \ldots & 0               \\
      \ldots    & \ldots  & \ldots & \ldots          \\
      0         & 0         & 0   & E_{2;d_P} \\
                       \end{array} \right). 
\end{eqnarray}
The unitary transformation matrix $U_a$ in which $a=2$ refers to two-body cluster 
approximation and can be split into four blocks corresponding to the blocks within the 
large squares of Fig.~\ref{Figure_Double_OLS}:
\begin{eqnarray}
\label{U_a_full}
 U_a = \left(\begin{array}{cc}
     U_a^P & U_a^{PQ} \\
     U_a^{QP} & U_a^Q \\
\end{array} \right), 
\end{eqnarray}
where the matrix $U_a^P$ is the $d_P\times d_P$ square matrix corresponding to the $P$ space.
One constructs the $U_2^P$ matrix from the $U_a$ matrix by taking $d_P$ rows and columns of 
the eigenvectors corresponding to the chosen $d_P$ eigenvalues:
\begin{eqnarray}
\label{U_a_P}
 U_2^P = \left(\begin{array}{cccc}
     b_{1,1}   & b_{1,2}   & \ldots & b_{1,d_P}    \\
     b_{2,1}   & b_{2,2}   & \dots  & b_{2,d_P}    \\
     \ldots    & \ldots    & \ldots & \ldots       \\
     b_{d_P,1} & b_{d_P,2} & \ldots & b_{d_P,d_P}  \\
\end{array} \right). 
\end{eqnarray}

The primary effective Hamiltonian $H_2^P$, signified by the box labeled ``$PH_{\rm eff}P$'' 
in Fig. \ref{Figure_Double_OLS}, can then be calculated by using the following formula:
\begin{eqnarray}
\label{H_diag_Nmax_eff}
 H_2^P & = &
            {U_2^{P\dagger} \over \sqrt{U_2^{P\dagger} U_2^P}}
            H^P_{2;\rm diag}
            {U_2^P \over \sqrt{U_2^{P\dagger} U_2^P}}  \nonumber\\
                       & = & T_{\rm rel} + V_{\rm eff}^P, 
\end{eqnarray}
where $V_{\rm eff}^P$ is the resulting primary effective $NN$ interaction and we suppress 
the subscript ``2.'' The interaction $V_{\rm eff}^P$ depends on $A$ and the chosen $P$ space 
including the selected value of $\Omega$. Note that the unitary transformation 
(\ref{H_diag_Nmax_eff}) is identical to OLS unitary transformation 
\cite{Okubo_1954,Suzuki_1980,Suzuki_1982} which satisfies the decoupling condition 
$QH_{\rm eff}P \defeq H_{2}^{QP}=0$ where the submatrix $QH_{\rm eff}P=0$ is one of two 
decoupling conditions depicted in Fig.~\ref{Figure_Double_OLS} for the primary Hamiltonian.

There are certain freedoms within the OLS renormalization procedure as well
as mathematical restrictions \cite{Viazminsky2001}. In this context, we note 
that in, our application, we select the $d_P$ lowest eigenvalues and eigenvectors
of $H_2$ for input to our primary effective Hamiltonian through Eq.~(\ref{H_diag_pmax}) 
and obtain numerically stable and accurate results.


\subsection{Transformation of many-body Hamiltonian into $sd$-shell space}

After a unitary transformation of the bare Hamiltonian in Eq.~ (\ref{H_2_body}) 
to the $4\hbar \Omega$ ($N_{\rm max}=4$) model space for 
the case of $^{18}$F, we calculate the $18$-body effective Hamiltonian 
 $PH_{\rm eff}P \defeq H_{18}^P$ in the $4\hbar \Omega$ space  
and solve for its low-lying eigenvalues and eigenvectors 
in a NCSM calculation.  This is analogous to solving the $a=2$ case
above so we introduce the corresponding subscript $18$.
We obtain a sufficient number of these 18-body solutions to generate
a second unitary transformation to take $H_{18}^{P}$ from the $4\hbar \Omega$ model 
space to a smaller secondary subspace $P'$, e.g., the $sd$-shell space, given 
by $N'_{\rm max}=0$. 
The secondary effective Hamiltonian is called $H_{a'}^{P'P}$ with $a'=18$ 
and is represented by $P'H'_{\rm eff}P'$ in Fig.~\ref{Figure_Double_OLS}).

This ``second step'' outlined above, follows a similar path to the ``first step''  
and is indicated by the work flow at the upper right in Fig.~\ref{Figure_Double_OLS}. 
Note that the $P$ space in the first unitary transformation is now split into parts 
related to the two subspaces, $P'$ and $Q'$, where $P' + Q' = P$. 
Our secondary effective Hamiltonian $H_{18}^{P'P}$ is designed to reproduce exactly 
the lowest $d_{P'}$ eigenvalues of the primary effective Hamiltonian $H^{P}_{18}$ through:
\begin{eqnarray}
\label{H_A_body_eff_sd}
H_{18}^{P'P} & = &
     {U_{18}^{P' \dagger} \over \sqrt{U_{18}^{P' \dagger} U_{18}^{P'}}}
            H^{P'}_{18;\rm diag}
            {U_{18}^{P'} \over \sqrt{U_{18}^{P' \dagger} U_{18}^{P'}}} \nonumber\\
                       & = & T_{\rm rel} + V_{\rm eff}^{P'P}, 
\end{eqnarray}
where $V_{\rm eff}^{P'P}$ is the resulting secondary effective interaction and we suppress 
the label for the $a'=18$ dependence.

This secondary effective Hamiltonian (\ref{H_A_body_eff_sd})  
is, in general, an $18$-body operator. However, in the  
$N'_{\rm max}=0$ case, the matrix dimension of the $18$-body secondary 
effective Hamiltonian  
(\ref{H_A_body_eff_sd}) is the same as the matrix dimension 
of a one-body plus two-body 
effective Hamiltonian acting in the $sd$-shell space. 
This means that $H_{18}^{P'P}$ can be taken to consist of only one-body 
and two-body terms, even after the exact $18$-body cluster transformation. 
All the orbitals below the $sd$-shell space are fully occupied by the other 
16 nucleon spectators, and the total $18$-body wave function can be exactly 
factorized into a $16$-body $0^+$ and two-body $sd$-shell wave functions. 
This considerably simplifies calculations with {$H_{18}^{P'P}$}.  
Therefore, we can write $a'$ as $a'=a_{\rm c}+a_{\rm v}$, 
where $a_{\rm c}$ is the number of core nucleons (16 in this case) 
and $a_{\rm v}$ is the size of the valence cluster.

In the third step outlined above we solve for the eigenvalues of $^{17}$F and $^{17}$O 
in the $P$ space using the effective interaction $V_{\rm eff}^P$ from 
Eq.~(\ref{H_diag_Nmax_eff}) joined with the $T_{\rm rel}$ for $A = 17$, 
to obtain the proton and neutron one-body terms of the secondary effective Hamiltonian
in the $sd$-shell space. Then we subtract the one-body terms from the secondary effective 
Hamiltonian of $^{18}$F, and we obtain the effective ``residual two-body interaction'' 
matrix elements (or simply the TBMEs) in the $sd$-shell space. 
Additionally, in the third step, we evaluate the $^{16}$O core energy by solving for its 
ground-state energy using the effective interaction  
$V_{\rm eff}^P$ from Eq.~(\ref{H_diag_Nmax_eff}) joined with $T_{\rm rel}$ for $^{16}$O.

Here, we adopt the $A = 16~(17)$ relative-kinetic-energy operators 
for NCSM evaluations of the core (single-particle) energies in step three.  
In earlier papers \cite{Lisetskiy_2008,Lisetskiy_2009}, 
based on the NCSM with a core first developed in Ref.~\cite{Navratil_1997}, 
a much  stronger $A$ dependence was obtained than in our present $sd$-shell calculations.  
We now understand these earlier results in terms of how the core 
and single-particle energies are calculated. In these earlier studies, 
the $A$ dependence of the kinetic-energy operator in the many-nucleon Hamiltonian 
used for calculating the core and single-particle energies was taken to be the total $A$ 
of the nucleus being studied.  In our current calculations, we use $A(\hbox{core}) = 16$ 
for the kinetic-energy operator when calculating the core energy 
and $A(\hbox{core}+1) = 17$ when calculating the single-particle energies, 
independent of the total $A$ of the nucleus being studied.  
Because the $A$ dependence of the kinetic energy operator goes as $1/A$, 
using the total $A$ instead of $A(\hbox{core})$ or $A(\hbox{core}+1)$ 
produces a much larger $A$ dependence of the core and single-particle energies.  
Both these choices are technically correct (i.e., they produce identical results for 
the nucleus being studied as we have verified), and merely reflect that these effective 
valence-space interactions are not uniquely defined.  
With our current choice, we achieve weak $A$ dependence of our resulting core, 
single-particle and valence effective interactions for $sd$-shell applications, 
which is appealing since this is a characteristic that is commonly found in phenomenological 
effective interactions.  Our weak $A$ dependence is also consistent with other {\it ab initio} 
investigations using either the IM-SRG technique \cite{Bogner_2014} or the coupled cluster 
method \cite{Jansen_2014}.

We then proceed to the fourth step and subtract this core energy 
from the energies of the single-particle states of $^{17}$F and $^{17}$O mentioned 
above to arrive at our valence single-particle energies. 
At the completion of step four, we have our twobody valence-core (2BVC)
effective Hamiltonian that may be used in standard shell-model (SSM) calculations. 

By using these core plus valence space single-particle energies along with the derived  
residual two-body effective interactions, we can perform the SSM calculations for $^{18}$F, 
as well as other nuclei, in the $sd$ shell and compare with full NCSM calculations 
in the $4\hbar \Omega$ space by using the primary effective Hamiltonian.
The SSM calculations for $^{18}$F will, by construction, 
give the same results as the NCSM calculations for $^{18}$F  
within numerical precision. 
A corresponding approach for $A > 18$ nuclei is exemplified below 
where we also provide a direct comparison between NCSM and SSM results.
One may then proceed, in principle, with SSM calculations to cases 
where full NCSM results are beyond current technical means.

We may summarize the results of steps 2-4 by arranging the results for the 
secondary effective Hamiltonian $H_{a'}^{P'P}$ into separate terms:
\begin{equation}
\label{H_ssm_two-body_eff}
{H}_{a'}^{P'P}=H_{a_c}^{P'P} + H_{\rm sp}^{P'P} + V_{a_v}^{P'P},
\end{equation}
where we have allowed for the more general case 
of two successive renormalization steps (signified by $P'P$) 
with $a' = A$ in the present discussion. In Eq.~(\ref{H_ssm_two-body_eff}) $H_{a_c}$ 
represents the core Hamiltonian for $a_c$ nucleons; $H_{\rm sp}$
represents the valence nucleon single-particle Hamiltonian and
$V_{a_v}$ represents the $a_v$-body residual effective valence interaction.  
Note that $V_{a_v}$ may be used for systems with more than 
$a_v$ valence nucleons, as we demonstrate below. 
We also note that the core and the valence single-particle Hamiltonians 
include their respective kinetic-energy terms.

In line with our approximations mentioned above, we use $^{18}$F alone to derive 
our isospin-dependent effective two-body interaction $V^{P'P}_{2}$ for the $sd$ shell. 
We then restrict our applications, at present, to cases with at most one proton in 
the $sd$ shell.

In SSM calculations, one typically uses
only the $H_{\rm sp}$ and $V_{a_v}$ terms in Eq.~(\ref{H_ssm_two-body_eff}).
In phenomenological Hamiltonians $H_{\rm sp}$ is often taken from experiment
and ${a_v}=2$ matrix elements are obtained by fits to properties 
of a set of nuclei.  We will present detailed comparisons between
our derived $H_{\rm sp}$ and $V_{a_v}$ terms with phenomenological 
interactions in a future presentation.

There is an important distinction between our SSM calculations
(with our Hamiltonian derived from the {\it ab initio} NCSM) and conventional
SSM calculations with phenomenological interactions.  We preserve the factorization
of the CM motion throughout our derivation for the primary and
secondary effective Hamiltonians.
Therefore, the $N'_{\rm max}=0$ secondary effective Hamiltonian not only reproduces
the appropriate $N_{\rm max}$ NCSM eigenvalues but also affords
access to wave functions for these $N'_{\rm max}=0$ states which may
be written with a factorized CM wave function of the entire system.

\section{EFFECTIVE TWO-BODY $sd$-SHELL INTERACTION}

In NCSM calculations, the dimension of the primary effective Hamiltonian 
increases very rapidly as we increase $ N_{\rm max}$ and/or the number of 
nucleons. We restricted the model space to $ N_{\rm max}=4$ in order to limit 
the computational effort, since our main goal is to demonstrate the procedure 
to obtain effective interactions in the $sd$ shell for the shell model with 
the $^{16}$O core using the \textit{ab initio} NCSM and to test these derived 
effective interactions with SSM calculations. In order to carry out NCSM 
calculations, we used the MFDn code \cite{Sternberg_2008,Maris_2010_2,Aktulga_2012} 
with the JISP16 and chiral N3LO $NN$ interactions. For the SSM calculations, 
we used a specialized version of the shell-model code ANTOINE 
\cite{Caurier_1999_1, Caurier_1999_2,Caurier_2001}.

\begin{table}[!t]
\caption{\label{tab:F18_Ex_ncsm_Nmax4}The NCSM energies (in MeV) of the lowest
28 states $J^\pi_i$ of $^{18}$F calculated in  $4\hbar \Omega$ model space by
using JISP16 and chiral N3LO $NN$ interactions with $\hbar \Omega=14$ MeV.}
\begin{ruledtabular}
\begin{tabular}{rrr|rrr}
$J^\pi_i$ & T & JISP16 & $J^\pi_i$ & T & N3LO \\ \hline
$1^+_1$ & 0  & $-122.742$ & $1^+_1$ & 0 & $-126.964$  \\
$3^+_1$ & 0  & $-122.055$ & $3^+_1$ & 0 & $-126.214$  \\
$0^+_1$ & 1  & $-121.320$ & $0^+_1$ & 1 & $-125.510$  \\
$5^+_1$ & 0  & $-120.329$ & $5^+_1$ & 0 & $-124.545$  \\
$2^+_1$ & 1  & $-119.505$ & $2^+_1$ & 1 & $-123.974$  \\
$2^+_2$ & 0  & $-119.011$ & $2^+_2$ & 0 & $-123.890$  \\
$1^+_2$ & 0  & $-118.709$ & $1^+_2$ & 0 & $-123.077$  \\
$0^+_2$ & 1  & $-118.410$ & $0^+_2$ & 1 & $-122.586$  \\
$2^+_3$ & 1  & $-117.211$ & $2^+_3$ & 1 & $-121.588$  \\
$3^+_2$ & 1  & $-117.035$ & $4^+_1$ & 1 & $-121.512$  \\
$4^+_1$ & 1  & $-117.004$ & $3^+_2$ & 1 & $-121.450$  \\
$3^+_3$ & 0  & $-116.765$ & $3^+_3$ & 0 & $-121.376$  \\
$1^+_3$ & 0  & $-113.565$ & $1^+_3$ & 0 & $-119.658$  \\
$4^+_2$ & 0  & $-112.314$ & $4^+_2$ & 0 & $-118.656$  \\
$2^+_4$ & 0  & $-111.899$ & $2^+_4$ & 0 & $-117.950$  \\
$1^+_4$ & 0  & $-110.357$ & $1^+_4$ & 0 & $-116.106$  \\
$4^+_3$ & 1  & $-109.625$ & $4^+_3$ & 1 & $-115.785$  \\
$2^+_5$ & 1  & $-109.292$ & $2^+_5$ & 1 & $-115.407$  \\
$1^+_5$ & 1  & $-108.752$ & $3^+_4$ & 0 & $-115.309$  \\
$3^+_4$ & 0  & $-108.706$ & $1^+_5$ & 1 & $-114.870$  \\
$2^+_6$ & 0  & $-108.485$ & $2^+_6$ & 0 & $-114.787$  \\
$1^+_6$ & 1  & $-108.055$ & $1^+_6$ & 1 & $-114.392$  \\
$2^+_7$ & 1  & $-108.041$ & $3^+_5$ & 1 & $-114.258$  \\
$3^+_5$ & 1  & $-107.874$ & $2^+_7$ & 1 & $-114.176$  \\
$3^+_6$ & 0  & $-101.528$ & $3^+_6$ & 0 & $-109.316$  \\
$1^+_7$ & 0  & $-99.946 $ & $1^+_7$ & 0 & $-107.798$  \\
$0^+_3$ & 1  & $-99.848 $ & $2^+_8$ & 1 & $-107.473$  \\
$2^+_8$ & 1  & $-99.607 $ & $0^+_3$ & 1 & $-107.436$  \\
\end{tabular}
\end{ruledtabular}
\end{table}

\subsection{Core and valence effective interactions for the $A=18$ system}

Following the methods presented in Sec. II for $H_2^P$ in Eq.~(\ref{H_diag_Nmax_eff}), 
we calculated the $18$-body primary effective Hamiltonians
with $N_{\rm max}=4$ and $\hbar \Omega=14$ MeV by using the bare JISP16
\cite{Shirokov_2007} and chiral N3LO \cite{Entem_2003} potentials for $V_{ij}^{NN}$.
We chose $\hbar \Omega=14$ MeV since it is near the minimum of the
ground-state energy of $^{16}$O at $N_{\rm max}=4$ \cite{Shirokov_2007} 
and it represents a typical choice for derived effective shell-model valence 
interactions (see, for example, Ref.~\cite{Vary:1977zz}). Future efforts with 
primary effective Hamiltonians derived in larger-$N_{\rm max}$ spaces will be 
needed for meaningful analyses of the $\hbar \Omega$ dependence of our results.

We solve for the $^{18}$F spectra in NCSM calculations with these primary effective 
Hamiltonians and present the lowest 28 eigenvalues in Table~\ref{tab:F18_Ex_ncsm_Nmax4}. 
The corresponding NCSM eigenvectors for these 28 states in the $N_{\rm max}=4$ space 
are the eigenvectors dominated by $N_{\rm max}=0$ components. These 28 eigenstates 
correspond with the complete set of $N_{\rm max}=0$ states in the $sd$ shell.

For each of these primary effective Hamiltonians $H_2^P$ we then followed steps 2-4 above 
to calculate secondary effective Hamiltonians $H_{18}^{P'P}$ as well as the resulting
six valence single-particle energies $H_{\rm sp}^{P'P}$ (three for neutrons and three for 
protons) and 63 valence two-body effective interaction matrix elements of $V_2^{P'P}$ in 
the coupled $JT$ representation. 

We now elaborate on the method of separating the secondary effective Hamiltonian 
$H_{18}^{P'P}$ into its components indicated in Eq.~(\ref{H_ssm_two-body_eff}).  
According to step 3 we first perform separate NCSM calculations for $^{17}$F and $^{17}$O 
using the Hamiltonian consisting of the same $V_{\rm eff}^{P}$ from 
Eq.~(\ref{H_diag_Nmax_eff}) combined with $T_{\rm rel}$ for $A=17$. These two calculations 
provide total single-particle energies for the valence protons and neutrons, respectively, 
that are expressed as matrix elements of $H_{a_c}^{P'P} + H_{\rm sp}^{P'P}$.

We continue with the second part of step 3 to obtain the  
core energy ($E_{\rm core}$) through a NCSM calculation for $^{16}$O by using 
the Hamiltonian consisting of $V_{\rm eff}^{P}$ from Eq.~(\ref{H_diag_Nmax_eff})
in combination with $T_{\rm rel}$ for $^{16}$O.
The resulting $^{16}$O ground-state energy defines the contribution of $H_{a_c}^{P'P}$ 
to the matrix elements of $H_{a_c}^{P'P} + H_{\rm sp}^{P'P}$ obtained in the 
$^{17}$F and $^{17}$O calculations. The valence single-particle energies, 
the eigenvalues of  $H_{\rm sp}^{P'P}$, are
then defined as the total single-particle energies less the core energy.

\begin{table} 
\caption{\label{tab:spe_JISP16} Proton and neutron single-particle energies (in MeV) 
for JISP16 effective interaction obtained for the mass of $A=18$ and $A=19$.}
\begin{ruledtabular}
\begin{tabular}{c|ccc|ccc}
 & \multicolumn{3}{c|}{$A=18$} & \multicolumn{3}{c}{$A=19$} \\ 
 & \multicolumn{3}{c|}{$E_{\rm core}=-115.529$} & \multicolumn{3}{c}{$E_{\rm core}=-115.319$} \\ \hline
$j_i$ & $\frac{1}{2}$ &$\frac{5}{2}$ & $\frac{3}{2}$ &  $\frac{1}{2}$ & $\frac{5}{2}$ & $\frac{3}{2}$  \\ \hline
$\epsilon_{j_i}^n$ & $-3.068$ & $-2.270$ & $6.262$ & $-3.044$ & $-2.248$ & $6.289$  \\
$\epsilon_{j_i}^p$ & $ 0.603$ & $ 1.398$ & $9.748$ & $ 0.627$ & $ 1.419$ & $9.774$   \\
\end{tabular}
\end{ruledtabular}
\end{table} 

\begin{table}
\caption{\label{tab:spe_N3LO} Proton and neutron single-particle energies (in MeV) 
for chiral N3LO effective interaction obtained for the mass of $A=18$ and $A=19$.}
\begin{ruledtabular}
\begin{tabular}{c|ccc|ccc}
 & \multicolumn{3}{c|}{$A=18$} & \multicolumn{3}{c}{$A=19$} \\
 & \multicolumn{3}{c|}{$E_{\rm core}=-118.469$} & \multicolumn{3}{c}{$E_{\rm core}=-118.306$} \\ \hline
$j_i$ & $\frac{1}{2}$ &$\frac{5}{2}$ & $\frac{3}{2}$ &  $\frac{1}{2}$ & $\frac{5}{2}$ & $\frac{3}{2}$   \\ \hline
$\epsilon_{j_i}^n$ & $-3.638$ & $-3.042$ & $3.763$ &  $-3.625$ & $-3.031$ & $3.770$   \\
$\epsilon_{j_i}^p$ & $ 0.044$ & $ 0.690$ & $7.299$ &  $ 0.057$ & $ 0.700$ & $7.307$   \\
\end{tabular}
\end{ruledtabular}
\end{table}

To obtain the TBMEs of the valence effective interaction $V_2^{P'P}$, we execute step 4 and 
subtract the contributions of the core and valence single-particle energies from the matrix 
elements of $H_{18}^{P'P}$ to isolate $V_2^{P'P}$ in Eq.~(\ref{H_ssm_two-body_eff}). 
To be specific, we designate our valence single-particle states by their angular momenta 
$j_i=\frac{1}{2}$, $\frac{3}{2}$, $\frac{5}{2}$. Then, we define the contribution
to the doubly reduced coupled-$JT$ TBMEs (signified by the subscript $JT$ on the TBME)
arising from the core and one-body terms as
\begin{multline}
\label{H_2BVC_onebody_tbme}
 \langle j_aj_b|| H_{a_c}^{P'P} +   H_{\rm sp}^{P'P}  ||j_cj_d\rangle_{JT}  \\
=
 (E_{\rm core} + \frac{1}{2}(\epsilon_{j_a}^n + \epsilon_{j_a}^p +  \epsilon_{j_b}^n  + \epsilon_{j_b}^p))
\delta_{j_a,j_c}\delta_{j_b,j_d}  ,
\end{multline}
where $\epsilon_j$ represents the valence single-particle energy for the orbital with 
angular momentum $j$ and the superscript, $n$~($p$), designates neutron (proton)
for the energy associated with the $^{17}$O ($^{17}$F) calculation, respectively.

The resulting doubly reduced coupled-$JT$ TBMEs of the valence effective interaction 
$V_2^{P'P}$ are expressed as
\begin{multline}
\label{H_2BVC_tbme}
 \langle j_aj_b|| V_{2}^{P'P} ||j_cj_d\rangle_{JT} \\ =  
\langle j_aj_b|| H_{a'}^{P'P} - H_{a_c}^{P'P} - H_{\rm sp}^{P'P}  ||j_cj_d\rangle_{JT}.
\end{multline}
By using the symmetries of the coupled-$JT$ representation, there are 63 unique
TBMEs for which $j_a \leq j_b$.  

We confirm the accuracy of this subtraction procedure by demonstrating that SSM calculations 
with the derived core, one-body, and two-body terms of Eq.~(\ref{H_ssm_two-body_eff}) in the 
$sd$-shell space reproduce the absolute energies of the lowest 28 states of the $4\hbar \Omega$
NCSM calculations for $^{18}$F shown in Table~\ref{tab:F18_Ex_ncsm_Nmax4}.

The results for the core energy ($E_{\rm core}$) and valence single-particle energies 
($\epsilon^n_j,\epsilon^p_j$) for the JISP16 interaction are presented on the left-hand side 
of Table~\ref{tab:spe_JISP16} for our leading example where the primary effective Hamiltonian 
is derived for $A=18$. The corresponding core energy and valence single-particle energy results 
for the chiral N3LO interaction are presented on the left-hand side of Table~\ref{tab:spe_N3LO}.  
The valence single-particle energies clearly reflect overall Coulomb energy shifts between 
NCSM calculations for $^{17}$F and $^{17}$O.

The resulting TBMEs of the secondary effective Hamiltonian $H_{18}^{P'P}$
in Eq.~(\ref{H_A_body_eff_sd}) and of the valence effective interaction $V_2^{P'P}$ 
in Eq.~(\ref{H_2BVC_tbme}) are given in the seventh and eighth columns respectively of
Tables~\ref{tab:H_2BVC_JISP16} and \ref{tab:H_2BVC_N3LO} in the Appendix.  
The results of Table~\ref{tab:H_2BVC_JISP16} are obtained with the JISP16 $NN$
interaction while those in Table~\ref{tab:H_2BVC_N3LO} are obtained with
the chiral N3LO $NN$ interaction. 

These results for $A=18$ with JISP16 presented in Tables~\ref{tab:spe_JISP16} and 
\ref{tab:H_2BVC_JISP16} (as well as the corresponding results with chiral N3LO in 
Tables~\ref{tab:spe_N3LO} and \ref{tab:H_2BVC_N3LO}) show the dominant contribution 
of $E_{\rm core}$ to the diagonal TBMEs of the secondary effective Hamiltonian 
$H_{18}^{P'P}$, as may be expected. When these $E_{\rm core}$ contributions along with 
the one-body contributions are subtracted following Eq.~(\ref{H_2BVC_tbme}), the resulting 
diagonal matrix elements of $V_2^{P'P}$ fall in the range of conventional phenomenological 
valence nucleon effective interactions. The nondiagonal TBMEs for $A=18$ shown in columns
seven and eight of Tables~\ref{tab:H_2BVC_JISP16} and \ref{tab:H_2BVC_N3LO} remain 
unchanged by the subtraction process of Eq.~(\ref{H_2BVC_tbme}) as required by the 
Kronecker deltas in Eq.~(\ref{H_2BVC_onebody_tbme}).

\begin{figure}[t]
\includegraphics[width= \columnwidth]{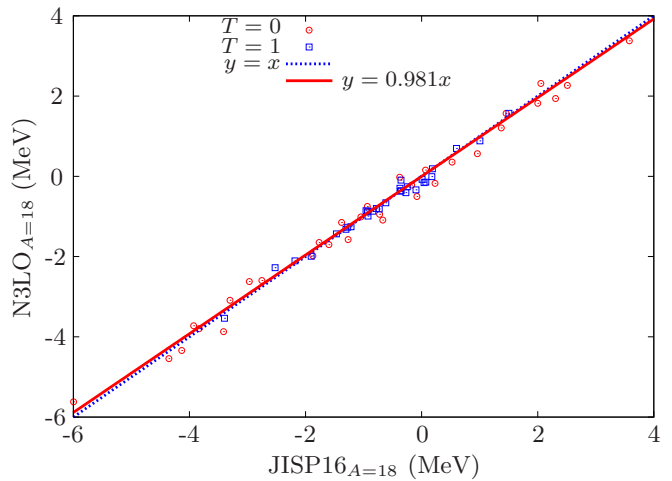}
\caption{(Color online) Correlation between chiral N3LO and JISP16 TBMEs plotted in units of MeV. 
The 63 TBMEs are derived for $A=18$ with the methods described in the text and are presented 
in the eighth columns of Tables~\ref{tab:H_2BVC_JISP16} and \ref{tab:H_2BVC_N3LO}. 
Red circles (blue squares) represent $T=0$~$(1)$ matrix elements. The diagonal dashed line is 
the reference for equal matrix elements. The solid red line is a linear fit to the correlation 
points with the result $y=0.981x$. The root-mean-square deviation between the two sets of TBMEs 
is 0.203 MeV. A plot of the $A=19$ results in the tenth columns of Tables~\ref{tab:H_2BVC_JISP16}
and \ref{tab:H_2BVC_N3LO} would be nearly indistinguishable from this plot.}
\label{fig:N3LO_A18_vs_JISP16_A18}
\end{figure}

The resulting TBMEs of $V_2^{P'P}$ in column eight of Tables~\ref{tab:H_2BVC_JISP16} and 
\ref{tab:H_2BVC_N3LO} (see tables in the Appendix) appear highly correlated, as shown in 
Fig.~\ref{fig:N3LO_A18_vs_JISP16_A18}, indicating significant independence of the valence
nucleon interactions from the underlying realistic $NN$ interaction. On the other hand, 
there is a noticeable dependence on the $NN$ interaction seen in the spin-orbit splitting 
of the valence single-particle energies in Tables~\ref{tab:spe_JISP16} and \ref{tab:spe_N3LO}.
For both the splitting of the $d_{5/2}$ and $d_{3/2}$ orbitals, and the splitting of the 
$s_{1/2}$ and the $d_{3/2}$ orbitals, the JISP16 interaction produces significantly larger 
results than the chiral N3LO interaction.  This is most noticeable in the approximately 30\%, 
or 2 MeV, larger splittings of the $d_{5/2}$ and $d_{3/2}$ orbitals obtained with JISP16. 

Note that both JISP16 and N3LO lead to splittings of the $d_{5/2}$ and $d_{3/2}$ orbitals 
that are larger than the phenomenological shell-model result which is based on experiment. 
In addition, the order of the calculated $s_{1/2}$ and $d_{5/2}$ orbitals are inverted 
compared with experiment. That is, $^{17}$F and $^{17}$O have a $5/2^+$ ground state, 
an excited $1/2^+$ at about 0.5 and 0.9 MeV respectively, and an excited $3/2^+$ at 
about 5 MeV. Of course, neither JISP16 nor the chiral N3LO interaction have been fit to 
any observable in the $sd$ shell. In addition, these calculated splittings should be 
sensitive to the 3NF, which is known to impact spin-orbit coupling effects in $p$-shell 
NCSM investigations \cite{Maris:2011as,Navratil_2007,Barrett_2013,Jurgenson_2013,Maris_2014}.

\subsection{Two-body valence cluster approximation for $A=19$ system}

We now illustrate our approach for going to heavier nuclei by adopting the specific 
example of $^{19}$F. In theory, we could proceed as with our application in the previous 
section, retain $a_c = 16$ and increase $a_v$ in pace with the increase with $A$.  
Thus, for $A=19$ we would derive matrix elements of an effective valence 3NF. However, 
this is not a practical path since there is no net gain over performing full NCSM 
calculations for each $A$ \textit{en route} to the secondary effective Hamiltonian.
Instead, we present an alternative approximate path to heavier nuclei.

Our procedure for going to heavier nuclei in the $sd$ shell is to specify the $sd$-shell 
nucleus of interest with its value of $A$ in the first step - the construction of 
the primary effective interaction $V_{\rm eff}^P$ of Eq.~(\ref{H_diag_Nmax_eff}).  
Then we define the two-body cluster Hamiltonian in Eq. (\ref{H_2_body}) 
with this new value of $A$ ($A=19$ in our specific example) which is 
subsequently used to construct the primary effective interaction.  
Next, we perform steps 2-4 as before with $a_c = 16$, $a_v = 2$, and neglecting 
effective many-valence-nucleon interactions: we perform $^{18}$F, $^{17}$F, $^{17}$O, 
and $^{16}$O NCSM calculations with this primary effective interaction $V_{\rm eff}^P$ 
in order to extract the core energy, proton and neutron valence single-particle energies, 
and valence TBMEs. This is the 2BVC applied for general $A$.
The generalization to $a_v = 3$ (the 3BVC approximation) is straightforward 
but computationally demanding. Note that for $A=19$ the 3BVC would correspond
to a complete NCSM calculation.

As an alternative, one may simply neglect any $A$ dependence of the core energy, 
valence single-particle energies, and valence TBMEs and perform SSM calculations 
throughout the $sd$ shell with the effective shell-model interaction derived for 
$^{18}$F. We also illustrate this choice below with the example of $^{19}$F.

We now investigate the consequences of neglecting the induced 3NF
and of neglecting the $A$ dependence of $V^P_{\rm eff}$.  That is,
we simply use the the derived core energy, valence single-particle energies, 
and valence TBMEs from the previous section in a SSM calculation of $^{19}$F.  
For comparison, we also derive these quantities specifically for the $^{19}$F system
in the 2BVC approximation, and we compare both with a complete NCSM calculations 
for $^{19}$F, which corresponds to performing the 3BVC approximation.

For the 2BVC approach to $^{19}$F, we perform step 1
beginning with $A=19$ instead of $A=18$ in Eqs.~(\ref{H_intrinsic})-(\ref{H_2_body}).  
That is, we calculate the primary effective Hamiltonian of Eq.~(\ref{H_diag_Nmax_eff})
for $^{19}$F instead of $^{18}$F. Then we proceed through the remaining equations,
as we did for $^{18}$F, using $V_{\rm eff}^P$ defined in Eq.~(\ref{H_diag_Nmax_eff}).
For example, in the second step we solve for the secondary effective Hamiltonian $H_{a'}^{P'P}$ 
with $a'=18$ at $N_{\rm max}=4$ using Eq.~(\ref{H_A_body_eff_sd}) as before. This 
establishes the foundation for proceeding with steps 3 and 4 to obtain the core energy, 
valence single-particle energies and valence TBMEs needed for solving 
$^{19}$F in a SSM calculation.

The resulting core energies and valence single-particle energies calculated by using 
JISP16 and chiral N3LO effective interactions are given in the right-hand columns of 
Tables~\ref{tab:spe_JISP16} and \ref{tab:spe_N3LO}, respectively. The core energies for 
the $A=19$ case are less attractive than the $A=18$ case by 210 keV (163 keV) for 
JISP16 (chiral N3LO). The single-particle energies for the $A=18$ and $A=19$ cases 
differ by less than 30 keV (20 keV) for JISP16 (chiral N3LO).  We observe, therefore, 
that the core and single-particle energies exhibit similarly weak $A$ dependence for 
both interactions.  

The resulting TBMEs of the secondary effective Hamiltonian $H_{18}^{P'P}$ in 
Eq.~(\ref{H_A_body_eff_sd}) and of the valence effective interaction $V_2^{P'P}$ 
in Eq.~(\ref{H_2BVC_tbme}) are given in the ninth and tenth columns, respectively, 
of Table \ref{tab:H_2BVC_JISP16} (for JISP16) and Table \ref{tab:H_2BVC_N3LO} 
(for chiral N3LO) in the Appendix. One observes a good correlation between the TBME 
results from the $A=18$ case and the $A=19$ case by comparing column seven with column nine 
and column eight with column ten in both Tables~\ref{tab:H_2BVC_JISP16} (for JISP16) 
and \ref{tab:H_2BVC_N3LO} (for chiral N3LO). The TBME's of $V_2^{P'P}$ exhibit particularly 
weak $A$ dependence. The largest difference between the TBMEs in column eight and column ten 
in Table~\ref{tab:H_2BVC_JISP16} (for JISP16) is 9 keV and corresponding largest difference 
in Table~\ref{tab:H_2BVC_N3LO} (for chiral N3LO) is 4 keV.

Our observed weak $A$ dependence of the core energies, valence single-particle
energies, and TBMEs is consistent with the view that the OLS transformation to the
$P$ space accounts for the high-momentum components of the $NN$ interaction 
and the results are approximately independent of whether the two-body cluster is treated as 
embedded in $A=18$ or in $A=19$. The similarity of the derived TBMEs is also suggestive 
of a common, or universal, soft effective $NN$ interaction.

We may elaborate on these points by noting that the first OLS transformation  
can be viewed as reducing the ultraviolet (UV) regulator of the JISP16 and N3LO 
interactions to the UV scale of the HO basis space limit controlled by $N_{\rm max}$ 
and by $\hbar\Omega$.  The HO basis space UV regulator imposed by our first OLS 
transformation may be estimated by using $N$, the maximum of $2n+l$ of the HO 
single-particle orbits included in the $P$ space. For $^{18}$F (or $^{19}$F) with 
$N_{\rm max}=4$ and $\hbar\Omega=14$ MeV this UV regulator is estimated to be either 
$\sqrt{(N+3/2)m\Omega} = 1.59$ fm$^{-1}$  \cite{Coon:2012ab} or 
$\sqrt{2(N+3/2+2)m\Omega} = 2.53$ fm$^{-1}$ \cite{More:2013rma}.  
In either case, the estimated UV regulator is independent of $A$ 
and is sufficiently low that we may speculate that our chosen $NN$ interactions 
are yielding a common (or universal) UV-regulated primary effective $NN$-interaction 
with the UV-regulation scale fixed by our choice of $P$ space.  Subsequent processing
through the second OLS transformation is the same for both primary effective $NN$ 
interactions so it retains that universality feature.

\subsection{SSM and NCSM calculations for $^{18}$F and  $^{19}$F with $A=18$
            and $A=19$ interactions}

We performed SSM calculations for the ground state and a few low-lying excited states 
of $^{18}$F and $^{19}$F by using the secondary effective Hamiltonians $H_{18}^{P'P}$ 
of Eq.~(\ref{H_ssm_two-body_eff}) developed from the JISP16 and chiral N3LO potentials.
We performed these SSM calculations with the code ANTOINE \cite{Caurier_1999_1, 
Caurier_1999_2,Caurier_2001} by explicitly summing the one-body and two-body
components on the right-hand side of Eq.~(\ref{H_ssm_two-body_eff}) whose matrix elements 
are presented in Tables~\ref{tab:spe_JISP16} and \ref{tab:H_2BVC_JISP16} for JISP16 and in 
Tables~\ref{tab:spe_N3LO}  and \ref{tab:H_2BVC_N3LO} for chiral N3LO. Next, we add the 
respective core energy to the resulting spectra to yield total energies for comparison 
with NCSM calculations performed with the primary effective Hamiltonian.

\begin{figure}[!t]
\includegraphics[width = \columnwidth]{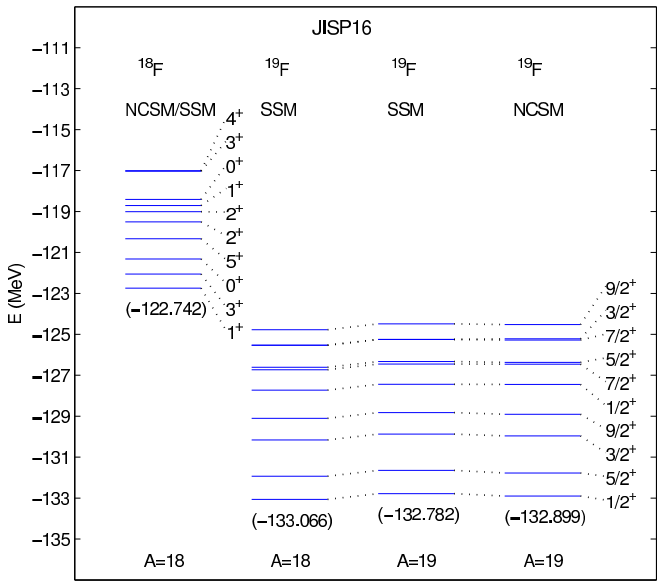}%
\caption{(Color online) The ground-state energy (in MeV) and low-lying excited-state 
energies of $^{18}$F and $^{19}$F obtained by the NCSM and SSM calculations using the 
effective JISP16 interaction. The tags $A=18$ and $A=19$ at the bottom of each column 
refer to the effective JISP16 interaction obtained with the 2BVC approximation for 
general $A$.  That is, the tags $A=18$ and $A=19$ represent nucleus $A$ used for deriving 
the primary effective Hamiltonian.  In addition, we retain only effective core, one-body, 
and two-body terms for the secondary effective Hamiltonian.}
\label{fig:ssm_ncsm_JISP16}
\end{figure}

\begin{figure}[t]
\includegraphics[width = \columnwidth]{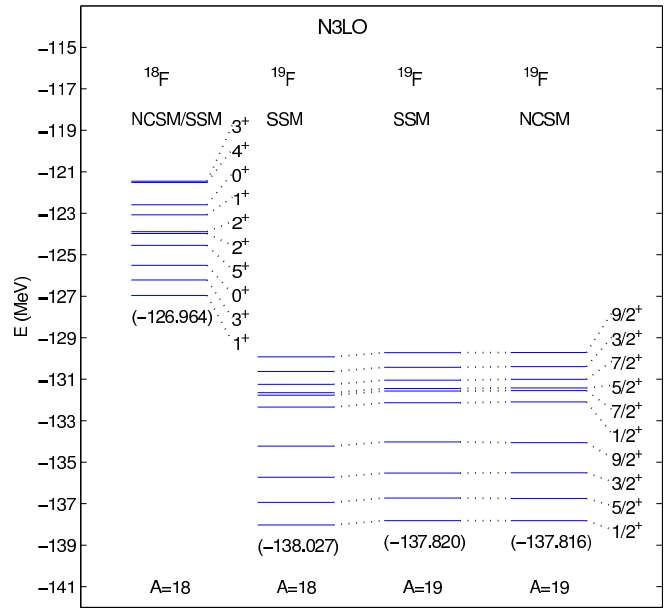}%
\caption{(Color online) The ground-state energy (in MeV) and low-lying excited-state 
energies of $^{18}$F and $^{19}$F obtained by the NCSM and SSM calculations using the 
effective chiral N3LO interaction. The tags $A=18$ and $A=19$ at the bottom of each 
column refer to the effective JISP16 interaction obtained with the 2BVC approximation 
for general $A$.  That is, the tags $A=18$ and $A=19$ represent nucleus $A$ used for 
deriving the primary effective Hamiltonian.  In addition, we retain only effective core, 
one-body, and two-body terms for the secondary effective Hamiltonian.}
\label{fig:ssm_ncsm_N3LO}
\end{figure}

We also carried out NCSM calculations for $^{18}$F and $^{19}$F by using 
the primary effective Hamiltonians $H_2^P$ of Eq.~(\ref{H_diag_Nmax_eff})
which are based on the selected $NN$ interaction and on the selected $A$ in
Eqs.~(\ref{H_intrinsic})--(\ref{H_2_body}). The SSM and NCSM results for the ground 
state and a few low-lying excited states of $^{18}$F and $^{19}$F are shown in 
Fig.~\ref{fig:ssm_ncsm_JISP16} for JISP16 and in Fig.~\ref{fig:ssm_ncsm_N3LO} 
for chiral N3LO. The nucleus for which the spectra are presented ($^{18}$F or $^{19}$F) 
is specified at the top of each column along with the many-body method - either NCSM 
with the primary effective Hamiltonian or SSM with the secondary effective Hamiltonian.  
Below each column we specify the $A$ used in Eqs.~(\ref{H_intrinsic})--(\ref{H_2_body}). 
When the results of the NCSM and SSM are the same with both many-body methods (as they 
should be theoretically for $^{18}$F), they appear as a single column with the label 
``NCSM/SSM.''  This situation, a simple cross check of the manipulations and the codes, 
is presented in the first column of Fig.~\ref{fig:ssm_ncsm_JISP16} for JISP16 and of 
Fig.~\ref{fig:ssm_ncsm_N3LO} for chiral N3LO.  Although these figures show only the lowest 
states, the cross-check is verified for all 28 states of two nucleons in the $sd$ shell.

 
The remaining three columns of Figs.~\ref{fig:ssm_ncsm_JISP16} and \ref{fig:ssm_ncsm_N3LO} 
display two SSM calculations with the secondary effective Hamiltonians and the
exact NCSM calculation, all for $^{19}$F. The second (third) column shows the results of 
using the primary effective Hamiltonian for $A=18$ ($A=19$) in the 2BVC approximation and 
solving the resulting SSM for $^{19}$F as outlined above. The difference between the second 
and third columns is interesting since it reflects two different 2BVC approximations.
In the second column, we see the effect of ignoring the contributions (both two body and 
three body) that one additional neutron makes by interacting with all nucleons in $^{18}$F. 
In the third column we see the effect of ignoring the contributions of all interactions 
in $^{19}$F to the effective three-body valence interaction in $^{19}$F.
The differences between columns two, three, and four (discussed further below)
are almost entirely due to the differences in the ground-state energies; the spectra are 
nearly the same. The ground-state energies in columns two and three in 
Figs.~\ref{fig:ssm_ncsm_JISP16}  and \ref{fig:ssm_ncsm_N3LO} differ over a range from 4 
to 211 keV compared with the exact results in column four.

The effects neglected in the two different approximations represented in columns two 
and three of Figs.~\ref{fig:ssm_ncsm_JISP16} and \ref{fig:ssm_ncsm_N3LO} led to small 
differences in the spectroscopy and, therefore, suggest that both are potentially 
fruitful paths for further investigation. However, when performing 2BVC calculations for 
$A>19$ nuclei (i.e., continuing to retain only core-, one- and two-body interaction terms) 
it is natural to expect that the difference between the SSM and NCSM calculations would 
increase due to the neglect of induced valence three-body, four-body, etc., interactions. 
The current results suggest that the dominant effect of neglecting these
higher-body induced interactions may appear mainly as an overall shift in the spectrum.
For the case of $^{19}$F, the shift between columns three and four in 
Fig.~\ref{fig:ssm_ncsm_JISP16} (\ref{fig:ssm_ncsm_N3LO}) shows that the 2BVC approximation 
for $A=19$ is responsible for an overall net attraction (repulsion) of about 117 keV (4~keV) 
which is small on the scale of the overall binding.  

The overall shift between columns two and three in Fig.~\ref{fig:ssm_ncsm_JISP16} 
(\ref{fig:ssm_ncsm_N3LO}) shows that the differences in our derived SSM Hamiltonians
produce about a 284 keV (207 keV) displacement in the binding energy. By referring 
to the results shown in Tables \ref{tab:spe_JISP16} and \ref{tab:spe_N3LO}, we find 
that this displacement in binding energies is attributed approximately to the difference 
in the core energies (about 80\% of the displacement) and to the difference in the sum of 
single-particle energies for the three valence nucleons (about 20\% of the displacement).
These displacements may be cast either as diagonal matrix elements of neglected induced 
3NFs or as corrections to the core and valence single-particle energies (or to a combination 
of both).  The distribution of these displacements will appear naturally when the 3BVC 
(i.e., full $a_v=3$) calculation is performed for $^{19}$F.

\section{SUMMARY, CONCLUSIONS, AND OUTLOOK}

We calculated $A$-dependent effective NCSM Hamiltonians, called primary effective Hamiltonians
[step 1 that leads to Eq.~(\ref{H_diag_Nmax_eff})], in a $4\hbar\Omega$ model space with the
realistic JISP16 and chiral N3LO $NN$ interactions. Next, we have solved the NCSM for low-lying
eigenstates sufficient to derive a secondary effective Hamiltonian that acts only in the
$0 \hbar\Omega$ model space for the $sd$ shell yet retains information from the full $A$-body
correlations present in the NCSM solutions [step 2 that leads to Eq.~(\ref{H_A_body_eff_sd})].
We then separate the TBMEs of the secondary effective Hamiltonians into core, one-body, and
two-body contributions [steps 3 and 4 that lead to Eq.~(\ref{H_ssm_two-body_eff})] which
defines to the 2BVC effective Hamiltonian suitable for SSM calculations. Finally, we use
these secondary effective Hamiltonians in SSM calculations and compare with exact NCSM
results based on the primary Hamiltonians for $A=18$ and $19$.

We estimate that the first OLS transformation on the JISP16
and chiral N3LO $NN$ interactions produces primary effective interactions
down to a sufficiently low UV regulator scale that we obtain a nearly common,
or universal, primary effective $NN$ interaction.  Subsequent processing
through the second OLS transformation retains universality features resulting
in TBMEs from JISP16 and chiral N3LO that are highly correlated as visualized
in Fig.~\ref{fig:N3LO_A18_vs_JISP16_A18}.

The SSM spectra for $A=18$ in the valence space are the same as the low-lying
NCSM spectra since our theory of the secondary effective Hamiltonian is derived
from the NCSM solutions obtained with the primary effective Hamiltonian.
With the 2BVC approximation, for which we present two approaches, there are
small differences between the SSM and NCSM spectra for the $^{19}$F system.
These differences are due to the omitted three-body effective interactions for the
$^{19}$F system and are observed primarily as overall shifts in the spectra
that are mainly due to shifts in the core energies. Close examination of the core,
one-body, and two-body components of the secondary effective Hamiltonians shows weak
$A$ dependence, which is encouraging for applications to heavier nuclei.

We will extend our investigations to obtain more complete results in $sd$ shell
by proceeding to a higher $N_{\rm max}$ model space for NCSM solutions with the
primary effective Hamiltonian. We will extend the 2BVC approximation to the 3BVC
approximation by including the three-body components of the secondary effective
Hamiltonians. In addition, we plan to incorporate initial 3NFs in the NCSM
calculations that complement the realistic $NN$ interactions.


\begin{acknowledgments}
This work was supported by Higher Education Council of Turkey (YOK),
by The Scientific and Technological Research Council of Turkey (TUBITAK-BIDEB),
by the US Department of Energy under Grants No. DESC0008485 (SciDAC/NUCLEI) and
No. DE-FG02-87ER40371, by the US National Science Foundation under Grants No.
PHYS-0845912 and No. 0904782, by the Ministry of Education and Science of the
Russian Federation under Contracts No. P521 and No. 14.V37.21.1297, and
by the Russian Foundation of Basic Research within the Project 15-02-06604.
Computational resources were provided by the National Energy Research Supercomputer
Center (NERSC), which is supported by the Office of Science of the U.S. Department
of Energy under Contract No. DE-AC02-05CH11231.
\end{acknowledgments}

\appendix 
\section{Tabulation of derived two-body matrix elements}
\label{app-TBMEs}

In this Appendix, we present the tables of our derived 2-body matrix elements (TBMEs).

\begin{table}[h]
\caption{The TBMEs (in MeV) of the secondary $sd$-shell effective Hamiltonian $H_{18}^{P'P}$
obtained from the NCSM calculation with $N_{\rm max}=4$, $\hbar \Omega=14$ MeV, 
and JISP16 potential for $^{18}$F are shown as well as the TBMEs of its residual 
valence effective interaction, $V_2^{P'P}$. Pairs of columns are labelled by the $A$ used in 
Eqs.~(\ref{H_intrinsic})--(\ref{H_2_body}) to develop the primary effective NCSM Hamiltonian 
as discussed in the text.}
\label{tab:H_2BVC_JISP16}
\begin{ruledtabular}
\begin{tabular}{rrrrrrrrrr}
 & & & & & & \multicolumn{2}{c}{$A=18$} & \multicolumn{2}{c}{$A=19$} \\
\cline{7-8} \cline{9-10}  \\
$2j_a$ & $2j_b$ & $2j_c$ & $2j_d$ & $J$ & $T$ &
$H_{18}^{P'P}$ &$V_2^{P'P}$ & $H_{18}^{P'P}$ & $V_2^{P'P}$ \\
\hline
 1 & 1 & 1 & 1 & 0 & 1 & $-120.176$ & $-2.182$ & $-119.917$ & $ -2.181$ \\
 1 & 1 & 3 & 3 & 0 & 1 & $  -0.924$ & $-0.924$ & $  -0.924$ & $ -0.924$ \\
 1 & 1 & 5 & 5 & 0 & 1 & $  -1.274$ & $-1.274$ & $  -1.274$ & $ -1.274$ \\
 3 & 3 & 3 & 3 & 0 & 1 & $-100.477$ & $-0.958$ & $-100.214$ & $ -0.958$ \\
 3 & 3 & 5 & 5 & 0 & 1 & $  -3.397$ & $-3.397$ & $  -3.396$ & $ -3.396$ \\
 5 & 5 & 5 & 5 & 0 & 1 & $-118.926$ & $-2.525$ & $-118.673$ & $ -2.525$ \\
 1 & 1 & 1 & 1 & 1 & 0 & $-121.296$ & $-3.302$ & $-121.032$ & $ -3.296$ \\
 1 & 1 & 1 & 3 & 1 & 0 & $  -0.378$ & $-0.378$ & $  -0.383$ & $ -0.383$ \\
 1 & 1 & 3 & 3 & 1 & 0 & $   0.231$ & $ 0.231$ & $   0.236$ & $  0.236$ \\
 1 & 1 & 3 & 5 & 1 & 0 & $   2.054$ & $ 2.054$ & $   2.052$ & $  2.052$ \\
 1 & 1 & 5 & 5 & 1 & 0 & $  -0.936$ & $-0.936$ & $  -0.939$ & $ -0.939$ \\
 1 & 3 & 1 & 3 & 1 & 0 & $-112.168$ & $-3.412$ & $-111.902$ & $ -3.406$ \\
 1 & 3 & 3 & 3 & 1 & 0 & $  -1.380$ & $-1.380$ & $  -1.384$ & $ -1.384$ \\
 1 & 3 & 3 & 5 & 1 & 0 & $   1.455$ & $ 1.455$ & $   1.456$ & $  1.456$ \\
 1 & 3 & 5 & 5 & 1 & 0 & $   0.525$ & $ 0.525$ & $   0.528$ & $  0.528$ \\
 3 & 3 & 3 & 3 & 1 & 0 & $-100.450$ & $-0.931$ & $-100.181$ & $ -0.925$ \\
 3 & 3 & 3 & 5 & 1 & 0 & $  -0.172$ & $-0.172$ & $  -0.173$ & $ -0.173$ \\
 3 & 3 & 5 & 5 & 1 & 0 & $   2.511$ & $ 2.511$ & $   2.508$ & $  2.508$ \\
 3 & 5 & 3 & 5 & 1 & 0 & $-113.957$ & $-5.997$ & $-113.698$ & $ -5.996$ \\
 3 & 5 & 5 & 5 & 1 & 0 & $   3.579$ & $ 3.579$ & $   3.580$ & $  3.580$ \\ 
  5 & 5 & 5 & 5 & 1 & 0 & $-117.448$ & $-1.047$ & $-117.191$ & $ -1.043$ \\
 1 & 3 & 1 & 3 & 1 & 1 & $-108.749$ & $ 0.007$ & $-108.487$ & $  0.009$ \\
 1 & 3 & 3 & 5 & 1 & 1 & $   0.042$ & $ 0.042$ & $   0.042$ & $  0.042$ \\
 \end{tabular}
\end{ruledtabular}\vspace{-.4ex}
\end{table}

\begin{widetext}
\small
\noindent\parbox[t]{.47\columnwidth}{
{TABLE IV: ({\em Continued.})\strut}
\begin{ruledtabular}
\begin{tabular}{rrrrrrrrrr}
 & & & & & & \multicolumn{2}{c}{$A=18$} & \multicolumn{2}{c}{$A=19$} \\
\cline{7-8} \cline{9-10}  \\
$2j_a$ & $2j_b$ & $2j_c$ & $2j_d$ & $J$ & $T$ &
$H_{18}^{P'P}$ &$V_2^{P'P}$ & $H_{18}^{P'P}$ & $V_2^{P'P}$ \\
\hline
 3 & 5 & 3 & 5 & 1 & 1 & $-108.057$ & $-0.097$ & $-107.798$ & $ -0.096$ \\
 1 & 3 & 1 & 3 & 2 & 0 & $-110.023$ & $-1.267$ & $-109.760$ & $ -1.264$ \\
 1 & 3 & 1 & 5 & 2 & 0 & $  -2.969$ & $-2.969$ & $  -2.968$ & $ -2.968$ \\
 1 & 3 & 3 & 5 & 2 & 0 & $  -1.873$ & $-1.873$ & $  -1.873$ & $ -1.873$ \\
 1 & 5 & 1 & 5 & 2 & 0 & $-117.279$ & $-0.081$ & $-117.021$ & $ -0.079$ \\
 1 & 5 & 3 & 5 & 2 & 0 & $  -1.597$ & $-1.597$ & $  -1.597$ & $ -1.597$ \\
 3 & 5 & 3 & 5 & 2 & 0 & $-112.093$ & $-4.133$ & $-111.826$ & $-4.124$ \\
 1 & 3 & 1 & 3 & 2 & 1 & $-109.374$ & $-0.618$ & $-109.113$ & $-0.617$ \\
 1 & 3 & 1 & 5 & 2 & 1 & $   1.504$ & $ 1.504$ & $   1.504$ & $ 1.504$ \\
 1 & 3 & 3 & 3 & 2 & 1 & $   0.185$ & $ 0.185$ & $   0.185$ & $ 0.185$ \\
 1 & 3 & 3 & 5 & 2 & 1 & $   0.601$ & $ 0.601$ & $   0.601$ & $ 0.601$ \\
 1 & 3 & 5 & 5 & 2 & 1 & $   1.005$ & $ 1.005$ & $   1.005$ & $ 1.005$ \\
 1 & 5 & 1 & 5 & 2 & 1 & $-118.667$ & $-1.469$ & $-118.411$ & $-1.469$ \\
 1 & 5 & 3 & 3 & 2 & 1 & $  -0.840$ & $-0.840$ & $  -0.840$ & $-0.840$ \\
 1 & 5 & 3 & 5 & 2 & 1 & $  -0.374$ & $-0.374$ & $  -0.374$ & $-0.374$ \\
 1 & 5 & 5 & 5 & 2 & 1 & $  -0.780$ & $-0.780$ & $  -0.780$ & $-0.780$ \\
 3 & 3 & 3 & 3 & 2 & 1 & $ -99.766$ & $-0.247$ & $ -99.503$ & $-0.247$ \\
 3 & 3 & 3 & 5 & 2 & 1 & $  -0.933$ & $-0.933$ & $  -0.933$ & $-0.933$ \\
 3 & 3 & 5 & 5 & 2 & 1 & $  -0.730$ & $-0.730$ & $  -0.730$ & $-0.730$ \\
 3 & 5 & 3 & 5 & 2 & 1 & $-108.232$ & $-0.272$ & $-107.973$ & $-0.271$ \\
 3 & 5 & 5 & 5 & 2 & 1 & $  -0.352$ & $-0.352$ & $  -0.352$ & $-0.352$ \\
 5 & 5 & 5 & 5 & 2 & 1 & $-117.617$ & $-1.216$ & $-117.364$ & $-1.216$ \\
 1 & 5 & 1 & 5 & 3 & 0 & $-121.030$ & $-3.832$ & $-120.770$ & $-3.828$ \\
 1 & 5 & 3 & 3 & 3 & 0 & $   0.068$ & $ 0.068$ & $   0.066$ & $ 0.066$ \\
 1 & 5 & 3 & 5 & 3 & 0 & $   1.373$ & $ 1.373$ & $   1.375$ & $ 1.375$ \\
 1 & 5 & 5 & 5 & 3 & 0 & $  -1.766$ & $-1.766$ & $  -1.768$ & $-1.768$ \\
 3 & 3 & 3 & 3 & 3 & 0 & $-102.271$ & $-2.752$ & $-102.006$ & $-2.750$ \\
 3 & 3 & 3 & 5 & 3 & 0 & $   2.000$ & $ 2.000$ & $   1.998$ & $ 1.998$ \\
 3 & 3 & 5 & 5 & 3 & 0 & $   0.961$ & $ 0.961$ & $   0.963$ & $ 0.963$ \\
 3 & 5 & 3 & 5 & 3 & 0 & $-108.629$ & $-0.669$ & $-108.367$ & $-0.665$ \\
 3 & 5 & 5 & 5 & 3 & 0 & $   2.308$ & $ 2.308$ & $   2.306$ & $ 2.306$ \\
 5 & 5 & 5 & 5 & 3 & 0 & $-117.125$ & $-0.724$ & $-116.870$ & $-0.722$ \\
 1 & 5 & 1 & 5 & 3 & 1 & $-117.022$ & $ 0.176$ & $-116.765$ & $ 0.177$ \\
 1 & 5 & 3 & 5 & 3 & 1 & $  -0.356$ & $-0.356$ & $  -0.356$ & $-0.356$ \\
 3 & 5 & 3 & 5 & 3 & 1 & $-107.888$ & $ 0.072$ & $-107.629$ & $ 0.073$ \\
 3 & 5 & 3 & 5 & 4 & 0 & $-112.314$ & $-4.354$ & $-112.049$ & $-4.347$ \\
 3 & 5 & 3 & 5 & 4 & 1 & $-109.863$ & $-1.903$ & $-109.605$ & $-1.903$ \\
 3 & 5 & 5 & 5 & 4 & 1 & $  -1.303$ & $-1.303$ & $  -1.303$ & $-1.303$ \\
 5 & 5 & 5 & 5 & 4 & 1 & $-116.766$ & $-0.365$ & $-116.513$ & $-0.365$ \\
 5 & 5 & 5 & 5 & 5 & 0 & $-120.329$ & $-3.928$ & $-120.075$ & $-3.927$ \\
\end{tabular}
\end{ruledtabular}}
%
%
\hfill\parbox[t]{.47\columnwidth}{
{TABLE V: The TBMEs (in MeV) of the secondary $sd$-shell effective Hamiltonian $H_{18}^{P'P}$
obtained from the NCSM calculation with $N_{\rm max}=4$, $\hbar \Omega=14$ MeV, 
and chiral N3LO potential for $^{18}$F are shown as well as the TBMEs of its residual 
valence effective interaction, $V_2^{P'P}$. Pairs of columns are labelled by the $A$ used in 
Eqs.~(\ref{H_intrinsic})--(\ref{H_2_body}) to develop the primary effective NCSM Hamiltonian 
as discussed in the text.\strut}
\label{tab:H_2BVC_N3LO}
\begin{ruledtabular}
\begin{tabular}{rrrrrrrrrr}
 & & & & & & \multicolumn{2}{c}{$A=18$} & \multicolumn{2}{c}{$A=19$} \\
\cline{7-8} \cline{9-10}  \\
$2j_a$ & $2j_b$ & $2j_c$ & $2j_d$ & $J$ & $T$ &
$H_{18}^{P'P}$ &$V_2^{P'P}$ & $H_{18}^{P'P}$ & $V_2^{P'P}$ \\
\hline
 1 & 1 & 1 & 1 & 0 & 1 & $-124.196$ & $-2.106$ & $-123.978$ & $-2.104$ \\
 1 & 1 & 3 & 3 & 0 & 1 & $  -0.991$ & $-0.991$ & $  -0.991$ & $-0.991$ \\
 1 & 1 & 5 & 5 & 0 & 1 & $  -1.268$ & $-1.268$ & $  -1.268$ & $-1.268$ \\
 3 & 3 & 3 & 3 & 0 & 1 & $-108.265$ & $-0.858$ & $-108.086$ & $-0.857$ \\
 3 & 3 & 5 & 5 & 0 & 1 & $  -3.538$ & $-3.538$ & $  -3.537$ & $-3.537$ \\
 5 & 5 & 5 & 5 & 0 & 1 & $-123.099$ & $-2.278$ & $-122.914$ & $-2.277$ \\
 1 & 1 & 1 & 1 & 1 & 0 & $-125.152$ & $-3.089$ & $-124.960$ & $-3.086$ \\
 1 & 1 & 1 & 3 & 1 & 0 & $  -0.022$ & $-0.022$ & $  -0.023$ & $-0.023$ \\
 1 & 1 & 3 & 3 & 1 & 0 & $  -0.175$ & $-0.175$ & $  -0.175$ & $-0.175$ \\
 1 & 1 & 3 & 5 & 1 & 0 & $   2.315$ & $ 2.315$ & $   2.314$ & $ 2.314$ \\
 1 & 1 & 5 & 5 & 1 & 0 & $  -0.750$ & $-0.750$ & $  -0.750$ & $-0.750$ \\
 1 & 3 & 1 & 3 & 1 & 0 & $-118.632$ & $-3.870$ & $-118.418$ & $-3.866$ \\
 1 & 3 & 3 & 3 & 1 & 0 & $  -1.149$ & $-1.149$ & $  -1.148$ & $-1.148$ \\
 1 & 3 & 3 & 5 & 1 & 0 & $   1.568$ & $ 1.568$ & $   1.568$ & $ 1.568$ \\
 1 & 3 & 5 & 5 & 1 & 0 & $   0.355$ & $ 0.355$ & $   0.355$ & $ 0.355$ \\
 3 & 3 & 3 & 3 & 1 & 0 & $-108.280$ & $-0.873$ & $-108.101$ & $-0.872$ \\
 3 & 3 & 3 & 5 & 1 & 0 & $  -0.217$ & $-0.217$ & $  -0.217$ & $-0.217$ \\
 3 & 3 & 5 & 5 & 1 & 0 & $   2.265$ & $ 2.265$ & $   2.264$ & $ 2.264$ \\
 3 & 5 & 3 & 5 & 1 & 0 & $-119.761$ & $-5.620$ & $-119.549$ & $-5.616$ \\
 3 & 5 & 5 & 5 & 1 & 0 & $   3.377$ & $ 3.377$ & $   3.375$ & $ 3.375$ \\
 5 & 5 & 5 & 5 & 1 & 0 & $-121.832$ & $-1.011$ & $-121.646$ & $-1.009$ \\
 1 & 3 & 1 & 3 & 1 & 1 & $-114.838$ & $-0.076$ & $-114.627$ & $-0.075$ \\
 1 & 3 & 3 & 5 & 1 & 1 & $  -0.157$ & $-0.157$ & $  -0.156$ & $-0.156$ \\
 3 & 5 & 3 & 5 & 1 & 1 & $-114.478$ & $-0.337$ & $-114.268$ & $-0.335$ \\
 1 & 3 & 1 & 3 & 2 & 0 & $-116.128$ & $-1.576$ & $-116.126$ & $-1.574$ \\
 1 & 3 & 1 & 5 & 2 & 0 & $  -2.623$ & $-2.623$ & $  -2.622$ & $-2.622$ \\
 1 & 3 & 3 & 5 & 2 & 0 & $  -1.980$ & $-1.980$ & $  -1.978$ & $-1.978$ \\
 1 & 5 & 1 & 5 & 2 & 0 & $-121.972$ & $-0.503$ & $-121.757$ & $-0.501$ \\
 1 & 5 & 3 & 5 & 2 & 0 & $  -1.703$ & $-1.703$ & $  -1.702$ & $-1.702$ \\
 3 & 5 & 3 & 5 & 2 & 0 & $-118.482$ & $-4.341$ & $-118.271$ & $-4.338$ \\
 1 & 3 & 1 & 3 & 2 & 1 & $-115.422$ & $-0.660$ & $-115.211$ & $-0.659$ \\
 1 & 3 & 1 & 5 & 2 & 1 & $   1.569$ & $ 1.569$ & $   1.569$ & $ 1.569$ \\
 1 & 3 & 3 & 3 & 2 & 1 & $   0.188$ & $ 0.188$ & $   0.188$ & $ 0.188$ \\
 1 & 3 & 3 & 5 & 2 & 1 & $   0.695$ & $ 0.695$ & $   0.695$ & $ 0.695$ \\
 1 & 3 & 5 & 5 & 2 & 1 & $   0.883$ & $ 0.883$ & $   0.883$ & $ 0.883$ \\
 1 & 5 & 1 & 5 & 2 & 1 & $-122.903$ & $-1.434$ & $-122.688$ & $-1.432$ \\
 1 & 5 & 3 & 3 & 2 & 1 & $  -0.869$ & $-0.869$ & $  -0.869$ & $-0.869$ \\
 1 & 5 & 3 & 5 & 2 & 1 & $  -0.298$ & $-0.298$ & $  -0.298$ & $-0.298$ \\
 1 & 5 & 5 & 5 & 2 & 1 & $  -0.802$ & $-0.802$ & $  -0.802$ & $-0.802$ \\
 3 & 3 & 3 & 3 & 2 & 1 & $-107.666$ & $-0.259$ & $-107.487$ & $-0.258$ \\
 3 & 3 & 3 & 5 & 2 & 1 & $  -0.885$ & $-0.885$ & $  -0.885$ & $-0.885$ \\
 \end{tabular}
\end{ruledtabular}}

\noindent\parbox{.47\columnwidth}{
{TABLE V: ({\em Continued.})\strut}
\begin{ruledtabular}
\begin{tabular}{rrrrrrrrrr}
 & & & & & & \multicolumn{2}{c}{$A=18$} & \multicolumn{2}{c}{$A=19$} \\
\cline{7-8} \cline{9-10}  \\
$2j_a$ & $2j_b$ & $2j_c$ & $2j_d$ & $J$ & $T$ &
$H_{18}^{P'P}$ &$V_2^{P'P}$ & $H_{18}^{P'P}$ & $V_2^{P'P}$ \\
\hline
 3 & 3 & 5 & 5 & 2 & 1 & $  -0.813$ & $-0.813$ & $  -0.813$ & $-0.813$ \\
 3 & 5 & 3 & 5 & 2 & 1 & $-114.549$ & $-0.408$ & $-114.340$ & $-0.407$ \\
 3 & 5 & 5 & 5 & 2 & 1 & $  -0.359$ & $-0.359$ & $  -0.359$ & $-0.359$ \\
 5 & 5 & 5 & 5 & 2 & 1 & $-122.077$ & $-1.256$ & $-121.892$ & $-1.255$ \\
 1 & 5 & 1 & 5 & 3 & 0 & $-125.266$ & $-3.797$ & $-125.049$ & $-3.793$ \\
 1 & 5 & 3 & 3 & 3 & 0 & $   0.155$ & $ 0.155$ & $   0.154$ & $ 0.154$ \\
 1 & 5 & 3 & 5 & 3 & 0 & $   1.206$ & $ 1.206$ & $   1.205$ & $ 1.205$ \\
 1 & 5 & 5 & 5 & 3 & 0 & $  -1.648$ & $-1.648$ & $  -1.647$ & $-1.647$ \\
 3 & 3 & 3 & 3 & 3 & 0 & $-110.003$ & $-2.596$ & $-109.822$ & $-2.593$ \\
 3 & 3 & 3 & 5 & 3 & 0 & $   1.819$ & $ 1.819$ & $   1.818$ & $ 1.818$ \\
 3 & 3 & 5 & 5 & 3 & 0 & $   0.564$ & $ 0.564$ & $   0.563$ & $ 0.563$ \\
 \end{tabular}
\end{ruledtabular}}
\addtocounter{table}{-1}
\hfil\parbox{.47\textwidth}{
{TABLE V: ({\em Continued.})\strut}
\begin{ruledtabular}
\begin{tabular}{rrrrrrrrrr}
 & & & & & & \multicolumn{2}{c}{$A=18$} & \multicolumn{2}{c}{$A=19$} \\
\cline{7-8} \cline{9-10}  \\
$2j_a$ & $2j_b$ & $2j_c$ & $2j_d$ & $J$ & $T$ &
$H_{18}^{P'P}$ &$V_2^{P'P}$ & $H_{18}^{P'P}$ & $V_2^{P'P}$ \\
\hline
 3 & 5 & 3 & 5 & 3 & 0 & $-115.233$ & $-1.092$ & $-115.023$ & $-1.090$ \\
 3 & 5 & 5 & 5 & 3 & 0 & $   1.940$ & $ 1.940$ & $   1.939$ & $ 1.939$ \\
 5 & 5 & 5 & 5 & 3 & 0 & $-121.768$ & $-0.947$ & $-121.583$ & $-0.946$ \\
 1 & 5 & 1 & 5 & 3 & 1 & $-121.476$ & $-0.007$ & $-121.262$ & $-0.006$ \\
 1 & 5 & 3 & 5 & 3 & 1 & $  -0.094$ & $-0.094$ & $  -0.094$ & $-0.094$ \\
 3 & 5 & 3 & 5 & 3 & 1 & $-114.287$ & $-0.146$ & $-114.078$ & $-0.145$ \\
 3 & 5 & 3 & 5 & 4 & 0 & $-118.684$ & $-4.543$ & $-118.472$ & $-4.539$ \\
 3 & 5 & 3 & 5 & 4 & 1 & $-116.134$ & $-1.993$ & $-115.924$ & $-1.991$ \\
 3 & 5 & 5 & 5 & 4 & 1 & $  -1.319$ & $-1.319$ & $  -1.318$ & $-1.318$ \\
 5 & 5 & 5 & 5 & 4 & 1 & $-121.190$ & $-0.369$ & $-121.006$ & $-0.369$ \\
 5 & 5 & 5 & 5 & 5 & 0 & $-124.545$ & $-3.724$ & $-124.358$ & $-3.721$ \\
\end{tabular}
\end{ruledtabular}}

\end{widetext}



\begin{thebibliography}{99} 

\bibitem{Stetcu_2005}I. Stetcu, B. R. Barrett, P. Navr\'atil, and J. P. Vary,
        Phys. Rev. C \textbf{71}, 044325 (2005).
\bibitem{Navratil_2005}P. Navr\'atil, W. E. Ormand, C. Forssen, and E. Caurier,
        Eur. Phys. J. A 25, 481 (2005).
\bibitem{Stetcu_2006}I.Stetcu, B.R. Barrett, P. Navr\'atil, and J.P. Vary,
        Phys. Rev. C \textbf{73}, 037307 (2006).
\bibitem{Nogga_2006}A. Nogga, P. Navr\'atil, B.R. Barrett, and J.P. Vary,
        Phys. Rev. C \textbf{73}, 064002 (2006).
\bibitem{Roth_2007} R. Roth and P. Navr\'atil, 
        Phys. Rev. Lett. \textbf{99}, 092501 (2007).
\bibitem{Maris_2009}P. Maris, J. P. Vary, and A. M. Shirokov,
         Phys. Rev. C \textbf{79}, 014308 (2009).
\bibitem{Maris_2010_1}P. Maris, A. M. Shirokov, and J. P. Vary,
         Phys. Rev. C \textbf{81}, 021301(R) (2010).
\bibitem{Maris:2011as}P.~Maris, J.~P.~Vary, P.~Navr\'atil, W.~E.~Ormand, H.~Nam and D.~J.~Dean,
         Phys.\ Rev.\ Lett.\  {\bf 106}, 202502 (2011).
\bibitem{Cockrell_2012}C. Cockrell, J. P. Vary, and P. Maris,
         Phys. Rev. C \textbf{86}, 034325 (2012).
\bibitem{Maris_2013}P. Maris and J. P. Vary,
         Int. J. Mod. Phys. E {\bf 22}, 1330016 (2013).
\bibitem{Navratil_1997}P. Navr\'atil, M. Thoresen, and B. R. Barrett,
        Phys. Rev. C \textbf{55}, R573 (1997).
\bibitem{Navratil_2000_1}P. Navr\'atil, J. P. Vary, and B. R. Barrett,
        Phys. Rev. Lett \textbf{84}, 5728 (2000).
\bibitem{Navratil_2000_2}P. Navr\'atil, J. P. Vary, and B. R. Barrett,
        Phys. Rev. C \textbf{62}, 054311 (2000).
\bibitem{Navratil_2001}P. Navr\'atil, J. P. Vary, W. E. Ormand, and B. R. Barrett,
        Phys. Rev. Lett. \textbf{87}, 172502 (2001).
\bibitem{Navratil_2007}P. Navr\'atil, V. G. Gueorguiev, J. P. Vary, W. E. Ormand,
         and A. Nogga, Phys. Rev. Lett. \textbf{99}, 042501 (2007).
\bibitem{Barrett_2013}B. R. Barrett, P. Navr\'atil, and J. P. Vary, 
         Prog. Part. Nucl. Phys. \textbf{69}, 131 (2013).
\bibitem{Jurgenson_2013}E. D. Jurgenson, P. Maris, R. J. Furnstahl, P. Navr\'atil,
         W. E. Ormand, and J. P. Vary, Phys. Rev. C \textbf{87}, 054312 (2013).
\bibitem{Shirokov_2014}A.~M.~Shirokov, V.~A.~Kulikov, P.~Maris, and J.~P.~Vary, 
         in $NN$ and $3N$ Interactions, edited by L.~D.~Blokhintsev and I.~I.~Strakovsky 
         (Nova Science, Hauppauge, NY, 2014), Chap. 8, p. 231,
         http:/\!/www.novapublishers.com/ catalog/product\verb+_+info.php?products\verb+_+id=49997.
\bibitem{Maris_2014}P. Maris, J.P. Vary, A. Calci, J. Langhammer, S. Binder and R. Roth, 
         Phys. Rev. C \textbf{90}, 014314 (2014).
\bibitem{Fujii_2007}S. Fujii, T. Mizusaki, T. Otsuka, T. Sebe, and A. Arima,
	Phys. Lett. B \textbf{650}, 9 (2007).
\bibitem{Lisetskiy_2008}A. F. Lisetskiy, B. R. Barrett, M. K. G. Kruse, 
        P. Navr\'atil, I. Stetcu and J. P. Vary, Phys. Rev. C \textbf{78} 044302 (2008).
\bibitem{Bogner_2014}S. K. Bogner, H. Hergert, J. D. Holt, A. Schwenk, 
         S. Binder, A. Calci, J. Langhammer, R. Roth, Phys. Rev. Lett. \textbf{113}, 142501 (2014)
\bibitem{Jansen_2014}G. R. Jansen, J. Engel, G. Hagen, P. Navr\'atil, A. Signoracci, 
         Phys. Rev. Lett. \textbf{113}, 142502 (2014)
\bibitem{Okubo_1954}S. Okubo, Prog. Theor. Phys.  \textbf{12}, 603 (1954).
\bibitem{Suzuki_1980}K. Suzuki and S. Y. Lee,
         Prog. Theor. Phys. \textbf{64}, 2091 (1980).
\bibitem{Suzuki_1982}K. Suzuki, Prog. Theor. Phys. \textbf{68}, 246 (1982);
         K. Suzuki and R. Okamoto, \textit{ibid} \textbf{70}, 439 (1983).
\bibitem{Shirokov_2007}A. M. Shirokov, J. P. Vary, A. I. Mazur, and T. A. Weber, 
         Phys. Lett. B \textbf{644}, 33 (2007); A Fortran code for JISP16 is available 
         at the website: nuclear.physics.iastate.edu.
\bibitem{Entem_2003}D. R. Entem, R. Machleidt, Phys. Rev. C \textbf{68} 041001 (2003).
\bibitem{Viazminsky2001}C.P. Viazminsky and J.P. Vary, J. Math. Phys., \textbf{42}, 2055 (2001).
\bibitem{Lisetskiy_2009}A. F. Lisetskiy, M. K. G. Kruse, B. R. Barrett, P.~Navr\'atil, 
         I. Stetcu and J.P. Vary, Phys. Rev. C \textbf{80}, 024315 (2009).
\bibitem{Sternberg_2008}P. Sternberg, E. G. Ng, C. Yang, P. Maris, J. P. Vary, 
         M. Sosonkina, and H. V. Le, \textit{Procedia of the 2008 ACM/IEEE Conference on 
         Supercomputing (SC 2008)},(IEEE Press, Piscataway, NJ, 2008), pp. 15:1-12.
\bibitem{Maris_2010_2}P. Maris, M. Sosonkina, J. P. Vary, E. G. Ng, C. Yang,
         \textit{Procedia Computer Science 1 (May 2010, ICCS 2010)}(Elsevier, Amsterdam, 2010), 
         pp. 97-106.
\bibitem{Aktulga_2012}H.M. Aktulga, C. Yang, E. G. Ng, P. Maris, and J. P. Vary,
         \textit{Lecture Notes in Computer Science 7484 (2012)}(Springer, Heidelberg, 2012), 
         pp. 830-842.
\bibitem{Caurier_1999_1}E. Caurier and F. Nowacki, Acta Phys. Pol. B \textbf{30}, 705, (1999).
\bibitem{Caurier_1999_2}E. Caurier, G. Martinez-Pinedo, F. Nowacki, A. Poves,
         J. Retamosa, and A. P. Zuker, Phys. Rev. C \textbf{59} 2033 (1999).
\bibitem{Caurier_2001}E. Caurier, P. Navr\'atil, W. E. Ormand, and J. P. Vary,
         Phys. Rev. C \textbf{64}, 051301(R) (2001).
\bibitem{Vary:1977zz}J.~P.~Vary and S.~N.~Yang, Phys.\ Rev.\ C {\bf 15}, 1545 (1977), 
         and references therein.
\bibitem{Coon:2012ab}S.~A.~Coon, M.~I.~Avetian, M.~K.~G.~Kruse, U.~van Kolck, P.~Maris and J.~P.~Vary,
         Phys.\ Rev.\ C {\bf 86}, 054002 (2012).
\bibitem{More:2013rma}S.~N.~More, A.~Ekstr\"om, R.~J.~Furnstahl, G.~Hagen and T.~Papenbrock,
         Phys.\ Rev.\ C {\bf 87}, 044326 (2013).
\end{thebibliography}
\end{document}